\documentclass{ws-ijbc}
\usepackage{graphicx}
\usepackage{epstopdf}
\begin{document}

%\catchline{}{}{}{}{} % Publisher's Area please ignore

\markboth{Jothimurugan {\it et al.,}}{Experimental evidence for vibrational resonance and enhanced signal transmission in Chua's circuits}

\title{EXPERIMENTAL EVIDENCE FOR VIBRATIONAL RESONANCE AND ENHANCED SIGNAL TRANSMISSION IN CHUA'S CIRCUIT}

\author{R.~JOTHIMURUGAN}

\address{Centre for Nonlinear Dynamics, School of Physics, \\ Bharathidasan University, Tiruchirappalli-620 024, \\ Tamilnadu, India\\
jothi.cnld@yahoo.com}

\author{K.~THAMILMARAN}
\address{Centre for Nonlinear Dynamics, School of Physics, \\ Bharathidasan University, Tiruchirappalli-620 024, \\ Tamilnadu, India\\
maran.cnld@gmail.com}

\author{S.~RAJASEKAR}
\address{School of Physics, \\ Bharathidasan University, Tiruchirappalli-620 024, \\ Tamilnadu, India\\
rajasekar@cnld.bdu.ac.in}

\author{M.~A.~F.~SANJU\'{A}N}
\address{Nonlinear Dynamics, Chaos and Complex Systems Group, Departamento de F\'{i}sica, \\ Universidad Rey Juan Carlos, Tulip\'{a}n s/n, 28933 M\'{o}stoles, \\ Madrid, Spain\\
miguel.sanjuan@urjc.es}
\maketitle

\begin{history}
\received{(to be inserted by publisher)}
\end{history}

\begin{abstract}
We consider a single Chua's circuit and a system of a unidirectionally coupled $n$-Chua's circuits driven by a biharmonic signal with two widely different frequencies $\omega$ and $\Omega$, where $\Omega \gg \omega$. We show experimental evidence for vibrational resonance in the single Chua's circuit and undamped signal propagation of a low-frequency signal in the system of $n$-coupled Chua's circuits where only the first circuit is driven by the biharmonic signal. In the single circuit, we illustrate the mechanism of vibrational resonance and the influence of the biharmonic signal parameters on the resonance. In the $n(=75)$-coupled Chua's circuits  enhanced propagation of low-frequency signal is found to occur for a wide range of values of the  amplitude of the high-frequency input signal and coupling parameter. The response amplitude of the $i$th circuit increases with $i$ and attains a saturation. Moreover, the unidirectional coupling is found to act as a low-pass filter. 
\end{abstract}


%\begin{multicols}{2}
\section{Introduction}

A nonlinear system driven by a biharmonic force with two widely different frequencies say, $\omega$ and $\Omega$ with $\Omega \gg \omega$, can exhibit resonance at the low-frequency $\omega$ when the amplitude $g$ or frequency $\Omega$ of the high-frequency component is varied. This high-frequency driving force induced resonance is termed as vibrational resonance \cite{ref1, ref2}. This phenomenon can be used to identify a weak signal as well as to enhance the response of a system. Vibrational resonance has been analysed in different kind of systems such as monostable \cite{ref5}, bistable \cite{ref1, ref6, ref7, ref2, ref8}, multistable \cite{ref9}, excitable \cite{ref10} and time-delayed \cite{ref11, ref12,ref13} systems. Experimental evidence of vibrational resonance in an analog circuit simulation of an excitable system \cite{ref7} and in a vertical cavity surface emitting laser system \cite{ref14,ref15} have been reported. Recently, its occurrence is studied in nonlinear maps \cite{ref16}, small-world networks \cite{ref18, ref17} and in coupled oscillators \cite{ref19}.\\   

Study of a nonlinear phenomenon in a variety of dynamical systems help us to deeply understand its various features and also explore its applicability in real practical situations. In this connection we wish to point out that an analysis of the different phenomena occurring in nonlinear electronic circuits is of great significance. In nonlinear circuit analysis, Chua's circuit is commonly used as a prototype circuit to investigate a variety of dynamics. Though the recent analysis on Chua's circuit and its variant circuits like the modified Chua's circuits, etc, is exhaustive, we mention some of the noteworthy analysis. Experimental study of jump resonance \cite{ref20}, infinite cascades of spirals and hubs \cite{ref21}, onset of Shilnikov chaos through mixed mode oscillations \cite{ref22}, amplitude death due to bidirectional coupling of conjugate variables \cite{ref23} and ghost resonance \cite{ref24} have been reported. Dynamical behavior and stability analysis of the memristor based fractional-order Chua's circuit \cite{ref25}, stability relation of synchronization in a network of Chua's circuits with time varying coupling \cite{ref26} and features of total sliding-mode control strategy leading to dynamics insensitive to variations in parameters and external disturbance \cite{ref27} are investigated. Applicability of a graph-theoretical approach on synchronization \cite{ref28}, adaptive synchronization developed for a general class of chaotic systems with unknown time-varying parameters and external perturbations \cite{ref29} and the chaotic synchronization for data assimilation \cite{ref30} have also been analysed in coupled Chua's circuits.\\

The effect of biharmonic force, particularly, the phenomenon of vibrational resonance has not yet been studied experimentally in the Chua's circuit. This is the goal of the present work. Specifically, we consider both a single Chua's circuit and a system of $n$-coupled Chua's circuits. We focus our interest on the experimental study of vibrational resonance in a single Chua's circuit and in a PSpice (Personal Simulation Program with Integrated Circuit Emphasis) circuit simulation of $n$-coupled Chua's circuits. Constructing a large size circuit on a circuit board and analysing its performance have limitations due to the physical effects such as parasitic capacitance effects, internal noise, and mismatch in the circuit components. Further, the tolerance effects and circuit loading can also affect the behavior of the circuit. Moreover, parametric identification during the initial stage of circuit design is difficult because for each parametric values one has to make a search of availability of the off-shelf components. In recent years, circuit simulators based on SPICE, for example PSpice, have been widely used for investigating the dynamics of circuits \cite{ref35, ref36, ref31, ref34, ref32, ref33}. Evaluation of circuit functions and performance through PSpice is more productive than on a breadboard. With PSpice one can quickly check a circuit idea and perform simulated test measurements and analysis which are difficult, inconvenient and unwise for the circuits built on a breadboard. Therefore, we preferred PSpice circuit simulation of a system of $n$-coupled Chua's circuits instead of its real hardware construction.\\

The single circuit system is driven by the periodic force $f \sin \omega t$ and $g\sin \Omega t$. The response of the circuit displays resonance when the parameter $g$ or $\Omega$ of the high-frequency force is varied. To characterize the vibrational resonance, we use the response amplitude $Q$, the ratio of amplitude $A_{\omega}$ of the output signal at the frequency $\omega$ of the input signal and the amplitude $f$ of the input periodic signal $f \sin\omega t$. $A_{\omega}$ can be measured from the power spectrum of the output of the circuit. The signature of the resonance is clearly seen in the quantity $Q$ and time series plot. We are able to identify the influences of the parameters $\omega$, $\Omega$ and $f$ on the vibrational resonance. The critical value of $g$ at which the resonance occurs increases linearly with the frequency $\Omega$ while the response amplitude at resonance decreases linearly with $\Omega$. In the case of a system of $n$-coupled Chua's circuits the coupling is unidirectional and only the first circuit is driven by the biharmonic force. The response amplitude $Q_i$ of the $i$th circuit either increases or decreases with $i$ depending upon the values of the control parameters and approaches a limiting value for large values of $i$. For a range of values of $g$ and the coupling constant (resistivity of the coupling resistor), an undamped signal propagation with the limiting value of $Q > Q_i$ is achieved. In this case for distant circuits the high-frequency component of the output signal is suppressed and the output signal becomes a rectangular pulse-like form.

\section{Single Chua's circuit}
\begin{figure}[ht]
\begin{center}
\includegraphics[width=0.45\linewidth]{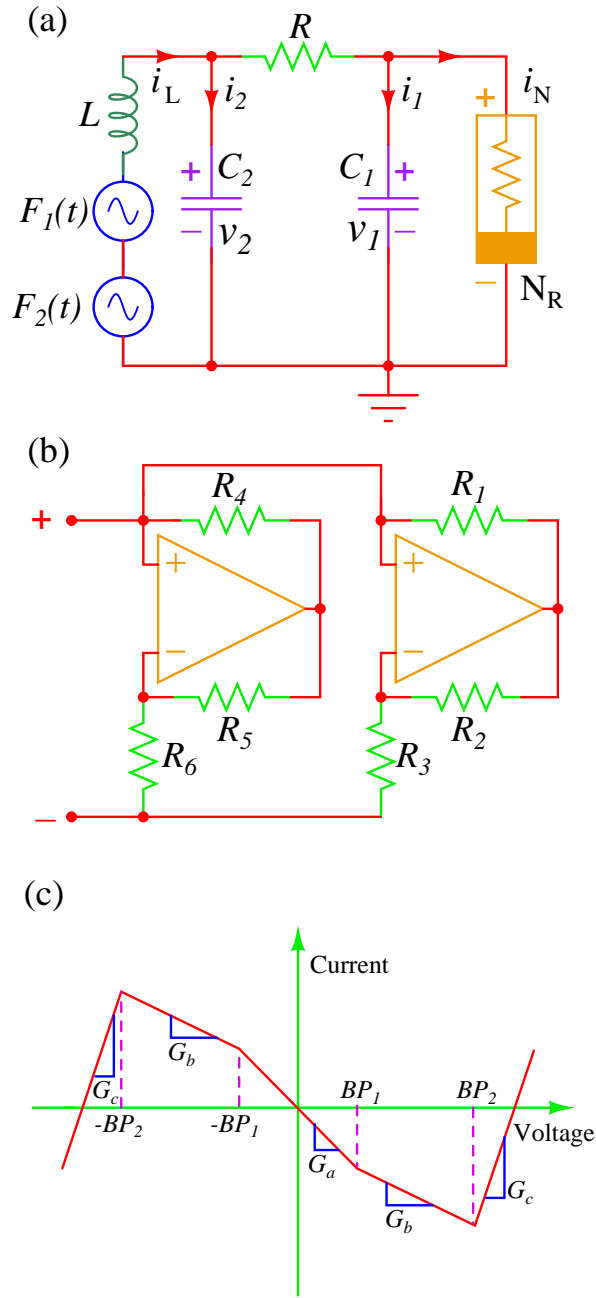}
\end{center}
\caption{(a) The Chua's circuit driven by two periodic forces, $F_1(t)=f \sin\omega t$ and $F_2(t)=g \sin\Omega t$. ${\mathrm{N_R}}$ represents the Chua's diode, the nonlinear element of the circuit. (b) Schematic diagram of the Chua's diode. Here $R_1=222.2~\Omega$, $R_2=224~\Omega$, $R_3=2.166~{\mathrm{k}}\Omega$, $R_4=21.63~{\mathrm{k}}\Omega$, $R_5=21.79~{\mathrm{k}}\Omega$ and $R_6=3.212~{\mathrm{k}}\Omega$. (c) The typical characteristic curve of the Chua's diode of (b).}
\label{vrcf1}
\end{figure}

In order to investigate the vibrational resonance, we drive the circuit by a biharmonic force of the form $F(t)=F_1(t) + F_2(t) = f \sin \omega t + g \sin \Omega t$ with $\Omega \gg \omega$. Figure \ref{vrcf1}(a) depicts the resultant Chua's circuit. The practical realization of the Chua's diode ${\mathrm{N_R}}$ is shown in Fig.~\ref{vrcf1}(b). ${\mathrm{N_R}}$ consists of two operational amplifiers and six linear resistors. The typical voltage-current characteristic of the Chua's diode is shown in Fig.~\ref{vrcf1}(c). It has five-segment piecewise linear form. Throughout  our study we fix the values of the circuit parameters as $C_1=10~{\mathrm{nF}}$, $C_2=100.5~{\mathrm{nF}}$, $L=18.75~{\mathrm{mH}}$, $R=1.98~{\mathrm{k}}\Omega$, $R_1=222.2~\Omega$, $R_2=224~\Omega$, $R_3=2.166~{\mathrm{k}}\Omega$, $R_4=21.63~{\mathrm{k}}\Omega$, $R_5=21.79~{\mathrm{k}}\Omega$ and $R_6=3.212~{\mathrm{k}}\Omega$. The values of the slopes in the voltage-current characteristic curve of the Chua's  diode are $G_a=-0.779~{\mathrm{mA/V}}$, $G_b=-0.4192~{\mathrm{mA/V}}$ and $G_c=4.464~{\mathrm{mA/V}}$, while the values of the break points are $BP_1=\pm 1.3489~{\mathrm{V}}$ and $BP_2=\pm 9.516~{\mathrm{V}}$. In the absence of a biharmonic force the Chua's circuit for the chosen parametric values has two stable equilibrium points $X_+=(v_1, v_2, i_L)=(5.656~{\mathrm{V}}, \,0~{\mathrm{V}}, \,-2.855~{\mathrm{mA}})$ and $X_-=(-5.656~{\mathrm{V}}, \,0~{\mathrm{V}}, \,2.855~{\mathrm{mA}})$ and one unstable equilibrium point $X_0=(0,\,0,\,0)$.\\

By applying Kirchhoff's laws to the various branches of circuit shown in Fig.1, we obtain the state equations as \cite{ref43,ref46,ref44}
\begin{subequations}
\label{eq1}
\begin{eqnarray}
C_1\frac{dv_1}{dt}&=&(1/R)(v_2-v_1)-f(v_1),\\
C_2\frac{dv_2}{dt}&=&(1/R)(v_1-v_2+i_L),\\
L\frac{di_L}{dt}&=&-v_2+f\sin\omega t+g\sin\Omega t,
\end{eqnarray}
\end{subequations} 
where $f(v_1)$ is the mathematical representation of Chua's diode characteristic curve: $f(v_1)=G_bv_1+0.5(G_a-G_b)[\lvert v_1+BP_1\rvert-\lvert v_1-BP_1\rvert]$. The dimensionless form of Eq.(\ref{eq1}) are
\begin{subequations}
\label{eq2}
\begin{eqnarray}
\dot{x}&=&\alpha(y-x-f(x)),\\
\dot{y}&=&x-y+z,\\
\dot{z}&=&\beta \left(-y+f'\sin\omega' \tau+g'\sin\Omega' \tau\right),
\end{eqnarray}
\end{subequations}   
where $f(x)=bx+0.5(a-b)[\lvert x+1 \rvert - \lvert x-1 \rvert]$, and $v_1=xBP_1$, $v_2=yBP_1$, $i_L=BP_1Gz$, $t=C_2\tau/G$, $G=1/R$, $a=RG_a$, $b=RG_b$, $\alpha=C_2/C_1$, $\beta=C_2R^2/L$, $f=f'BP_1$, $g=g'BP_1$, $\omega=\omega'C_2/G$ and $\Omega=\Omega'C_2/G$.\\

%%%%%%%%%%%%%%%%%%%%%%%%%%%%%%%%%%%%%%%%%%%%%%%%%%%%%%%%%%%%%%%%%%%%%
%%%%%%%%%%%%%%%%%%%%%%%%%%%%%%%%%%%%%%%%%%%%%%%%%%%%%%%%%%%%%%%%%%%%% 

In the experiment we consider that $\Omega \gg \omega$. Because the driving input signal has two  widely different frequencies $\omega$ (low-frequency) and $\Omega$ (high-frequency) the output signal of the circuit has components at these two frequencies and their linear combinations. Assume that in the absence of high-frequency input signal $g \sin \Omega t$, the amplitude $A_{\omega}$ of the output signal at the low-frequency $\omega$ is weak. We are interested in enhancing the amplitude $A_{\omega}$ of the output signal at the frequency $\omega$ by the high-frequency input signal $g \sin \Omega t$. To measure $A_\omega$ we consider the fast Fourier transform (FFT) of the output signal obtained using the Agilent (MSO6014A) Mixed Signal Oscilloscope. A small fluctuation of $A_{\omega}$ is observed in the FFT displayed in the instrument. In view of this for better accuracy an average value of $A_{\omega}$ over $25$ measurements is obtained. The value of $A_{\omega}$ measured in the FFT is in ${\mathrm{{dBV}}}$. It is then converted into the units of ${\mathrm{V}}$. Then we compute $Q=A_{\omega}$(in ${\mathrm{V}}$)$/f$ and is termed as the response amplitude of the circuit at the frequency $\omega$ \cite{ref1,ref2}.\\

\begin{figure}[ht]
\begin{center}
\includegraphics[width=0.6\linewidth]{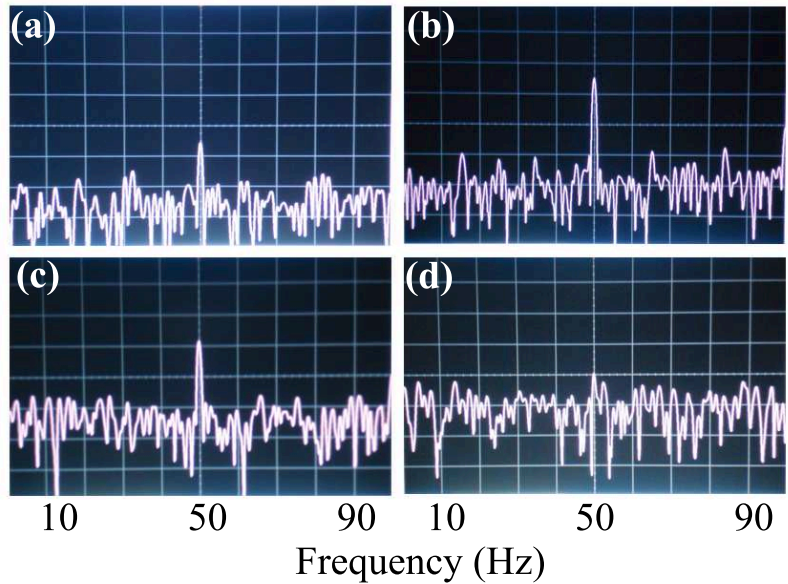}
\end{center}
\caption{The power spectrum of the voltage $v_1$ for four values of $g$. (a) $g=1~{\mathrm{V}}$, (b) $g=1.3~{\mathrm{V}}$, (c) $g=1.55~{\mathrm{V}}$ and (d) $g=2~{\mathrm{V}}$. The values of other circuit parameters are $C_1=10~{\mathrm{nF}}$, $C_2=100~{\mathrm{nF}}$, $L=18~{\mathrm{mH}}$, $R=1.98~{\mathrm{k}}\Omega$, $A=0.3~{\mathrm{V}}$, $\omega=50~{\mathrm{Hz}}$ and $\Omega=500~{\mathrm{Hz}}$.}
\label{vrcf2}
\end{figure}

\begin{figure}[t]
\begin{center}
\includegraphics[width=0.38\linewidth]{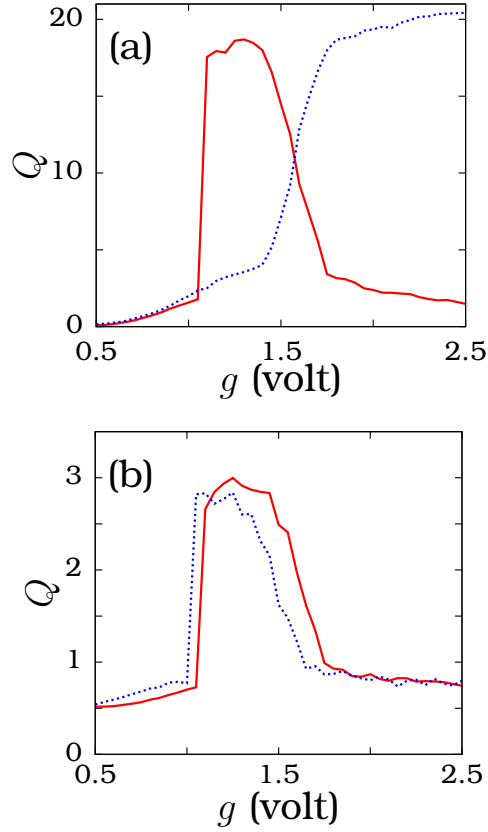}
\end{center}
\caption{(a) Response amplitudes $Q$ at the low-frequency $\omega$ (continuous curve) and high-frequency $\Omega$ (dotted curve) associated with $v_1$ versus the control parameter $g$. (b) $Q$ at the low-frequency $\omega$ associated with $v_2$ (continuous curve) and $i_{\mathrm{L}}$ (dotted curve) versus $g$.}
\label{vrcf3}
\end{figure}

We fix $f=0.3~{\mathrm{V}}$, $\omega=50~{\mathrm{Hz}}$ and $\Omega=500~{\mathrm{Hz}}$. Figure \ref{vrcf2} displays the power spectrum of the voltage $v_1$ of the circuit for four fixed values of the amplitude $g$ of the high-frequency input signal. The amplitude $A_\omega$ at $\omega=50~{\mathrm{Hz}}$ increases and then decreases which is a typical signature of resonance. To characterize the resonance we compute $Q$ at the frequencies $\omega$ and $\Omega$ of the voltages $v_1$ and $v_2$ and the current $i_L$ for a range of values of the control parameter $g$. Figure~\ref{vrcf3}(a) shows the variation of $Q$ of $v_1$ at $\omega$ and $\Omega$ with $g$. As $g$ increases from a small value $Q$ at $\omega$ increases slowly, varies sharply over an interval and reaches a maximum  value at a critical value of $g$ denoted as $g_{_{\mathrm{VR}}}$. The value of $g_{_{\mathrm{VR}}}$ is found as $1.3~{\mathrm{V}}$. When $g$ is increased further from $g_{_{\mathrm{VR}}}$ the response amplitude decreases rapidly to a small value. This type of resonance behavior is not observed with $Q$ at $\Omega$. In Fig.~\ref{vrcf3}(b) we plot $Q$ of $v_2$ and $i_L$ at $\omega$ as a function of $g$. Both $Q$ at $v_2$ and $i_L$ display resonance. In the rest of the analysis on the single Chua's circuit we consider $Q$ of $v_1$ at $\omega$.
%%%%%%%%%%%%%%%%%%%%%%%%%%%%%%%%%%%%%%%%%%%%%%%%%%%%%%%%%%%%%%%%%%%%%
%%%%%%%%%%%%%%%%%%%%%%%%%%%%%%%%%%%%%%%%%%%%%%%%%%%%%%%%%%%%%%%%%%%%%
%%%%%%%%%%%%%%%%%%%%%%%%%%%%%%%%%%%%%%%%%%%%%%%%%%%%%%%%%%%%%%%%%%%%%
%%%%%%%%%%%%%%%%%%%%%%%%%%%%%%%%%%%%%%%%%%%%%%%%%%%%%%%%%%%%%%%%%%%%%

\begin{figure}[th]
\begin{center}
\includegraphics[width=0.61\linewidth]{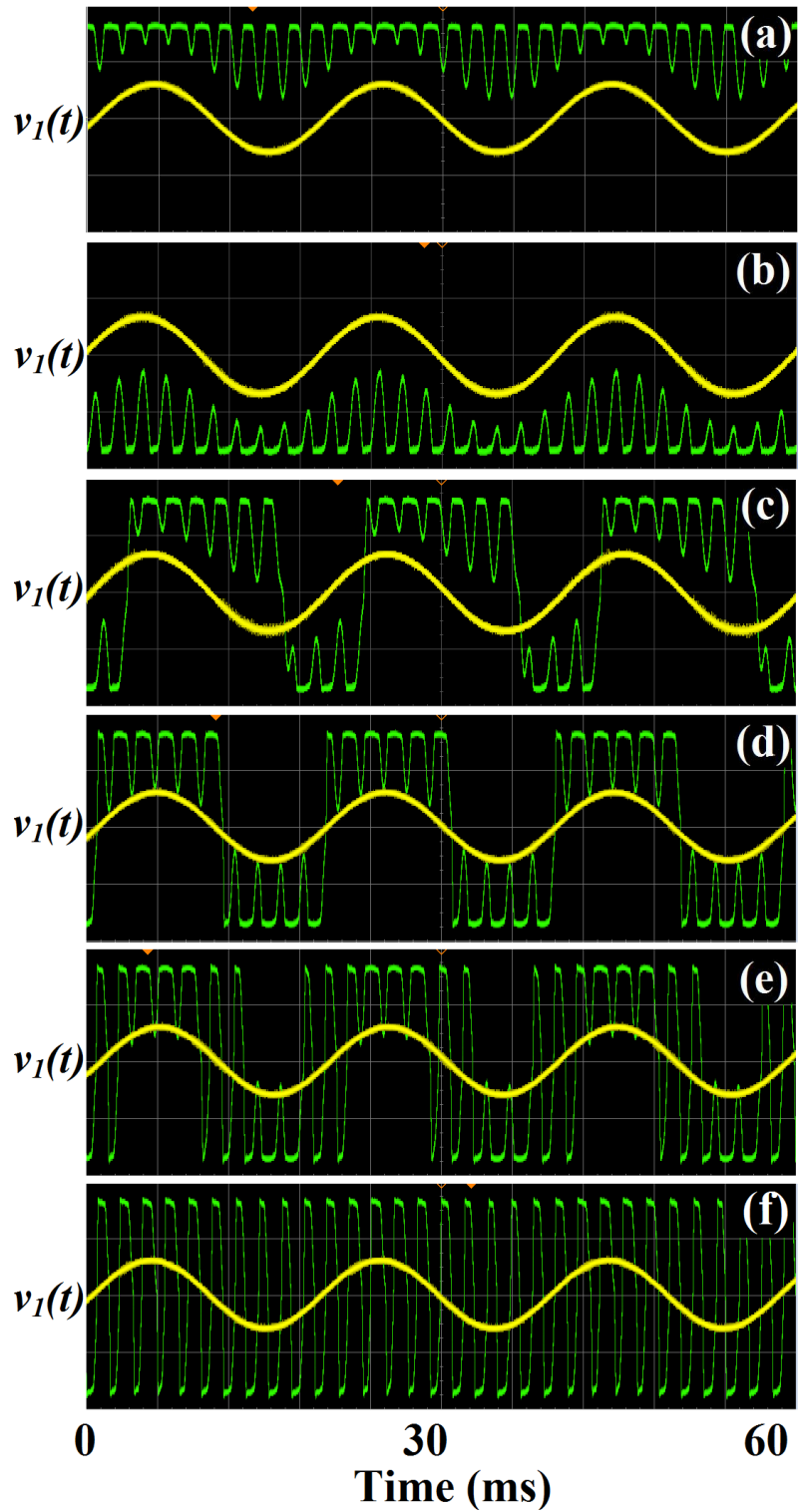}
\end{center}
\caption{Plot of $v_1(t)$ (green line) and the low-frequency periodic input signal $f\sin\omega t$ (yellow line) versus time $t$ for five values of the amplitude $g$ of the high-frequency periodic signal. (a) and (b) $g=1~{\mathrm{V}}$, (c) $g=1.15~{\mathrm{V}}$, (d) $g=1.3~{\mathrm{V}}$, (e) $g=1.55~{\mathrm{V}}$ and (f) $g=2~{\mathrm{V}}$. In all the subplots the range of $v_1(t)$ is [$-10~{\mathrm{V}}$,\,$10~{\mathrm{V}}$]. The low-frequency periodic input is $10$ times enlarged in all the subplots for clarity.}
\label{vrcf4}
\end{figure}

Next, we illustrate the mechanism of vibrational resonance using a time series plot. Figure~\ref{vrcf4} presents $v_1(t)$ versus $t$ for six fixed values of $g$. In the absence of a biharmonic force the Chua's circuit for the chosen parametric values has two stable equilibrium points $X_+=(v_1, v_2, i_L)=(5.656~{\mathrm{V}}, \,0~{\mathrm{V}}, \,-2.855~{\mathrm{mA}})$ and  {\noindent $X_-=(-5.656~{\mathrm{V}}, \,0~{\mathrm{V}}, \,2.855~{\mathrm{mA}})$ and one unstable equilibrium point $X_0=(0,\,0,\,0)$. For $g=0~{\mathrm{V}}$, $f=0.3~{\mathrm{V}}$ and $\omega=50~{\mathrm{Hz}}$, two period-$T$ ($=1/\omega$) orbits coexist - one orbit about $X_+$ and another about $X_-$. That is, on either side of $v_1=0$. When the system is driven further by the high-frequency force with $\Omega=500~{\mathrm{Hz}}$ then for small values of $g$, the two periodic orbits coexist and $v_1(t)$ is modulated by the high-frequency force. This is shown in Fig.~\ref{vrcf4}(a) and (b) for $g=1~{\mathrm{V}}$. At a certain value of $g$ crossing of $v_1=0$ takes place. We denote $\tau^+$ as the time spent by the trajectory in the region $v_1 > 0$ before switching to the region $v_1 < 0$. Similarly, we define $\tau^-$. $\tau^+$ and $\tau^-$ are the residence times of the trajectory in the regions $v_1 > 0$ and $v_1 < 0$ respectively. We can then calculate the mean residence times $\tau^+_{_{\mathrm{MR}}}$ and $\tau^-_{_{\mathrm{MR}}}$. For $g$ values just above the onset of switching between $v_1 < 0$ and $v_1 > 0$, the residence times $\tau^+_{_{\mathrm{MR}}}$ and $\tau^-_{_{\mathrm{MR}}}$ are unequal. An example is shown in Fig.~\ref{vrcf4}(c) where $g=1.15~{\mathrm{V}}$. $\tau^+_{_{\mathrm{MR}}}$ and $\tau^-_{_{\mathrm{MR}}}$ vary with $g$. At the critical value $g=g_{_{\mathrm{VR}}}=1.3~{\mathrm{V}}$, $\tau^+_{_{\mathrm{MR}}}=\tau^-_{_{\mathrm{MR}}}=T/2$ (see Fig.~\ref{vrcf4}(d)). There is a periodic switching between the regions $v_1 < 0$ and $v_1 > 0$ with period equal to half the time of the period of the low-frequency input signal. The response amplitude $Q$ is maximum at this value of $g$. This is the mechanism for vibrational resonance. Note that $Q$ is not maximum, that is resonance does not occur, at the value of $g$ for which the onset of crossing occurs. When $g$ is further increased from $g_{_{\mathrm{VR}}}$ the synchronization between $v_1(t)$ and the input signal $f \sin\omega t$ is lost (see Fig.~\ref{vrcf4}(e)). For sufficiently large values of $g$ a rapid switching between the regions $v_1 < 0$ and $v_1 > 0$ occurs and now the oscillation is centered around the equilibrium point $X_0$. This is evident in Fig.~\ref{vrcf4}(f) where $g=2~{\mathrm{V}}$.}\\

\begin{figure}[th]
\begin{center}
\includegraphics[width=0.65\linewidth]{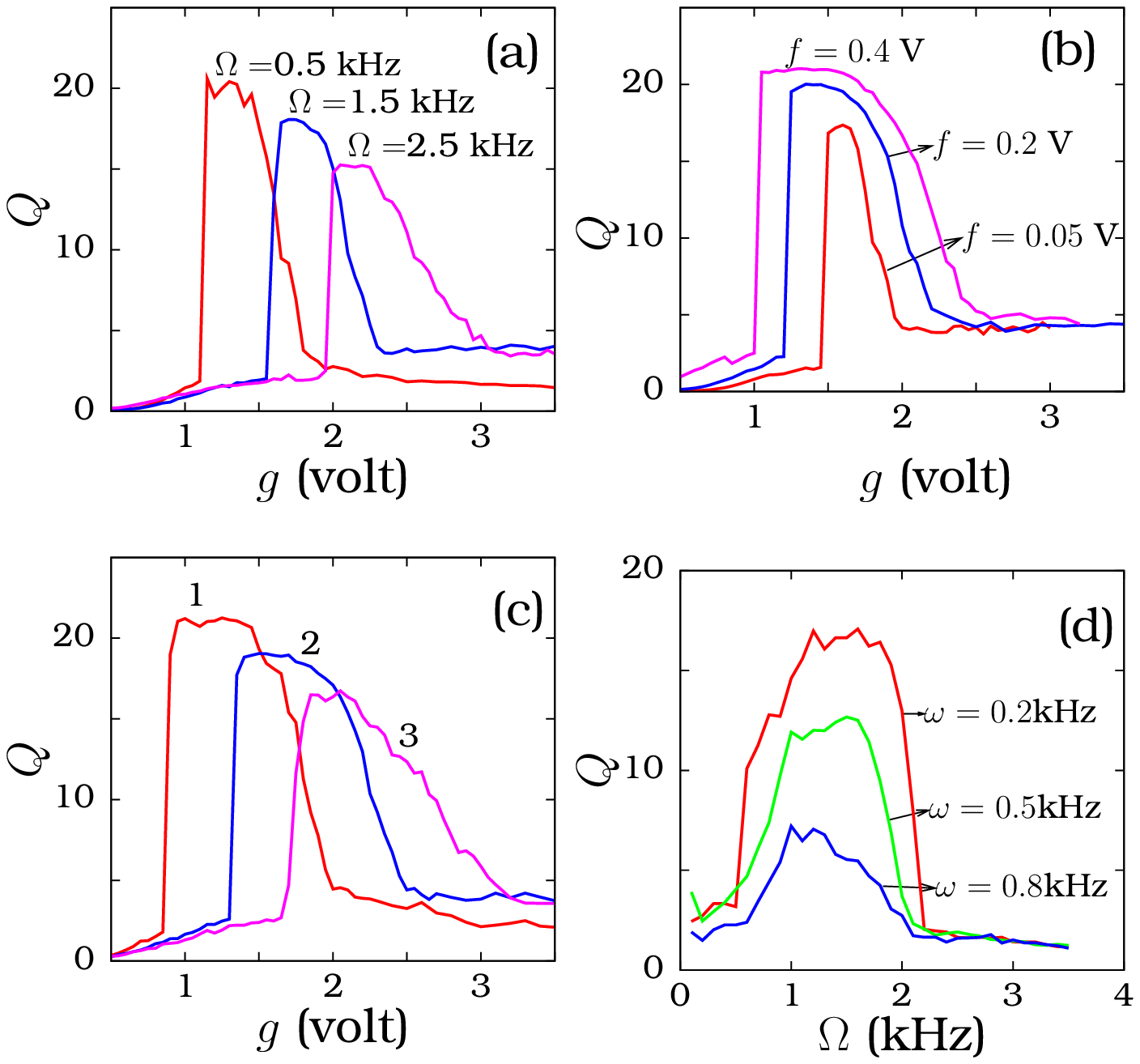}
\end{center}
\caption{The dependence of the response amplitude ($Q$) of the voltage $v_1$ versus the parameter $g$ on (a) the high-frequency $\Omega$ = $0.5~{\mathrm{kHz}}$, $1.5~{\mathrm{kHz}}$ and $2.5~{\mathrm{kHz}}$ with $f=0.3~{\mathrm{V}}$ and $\omega=50~{\mathrm{Hz}}$, (b) the amplitude $f$ = $0.05~{\mathrm{V}}$, $0.2~{\mathrm{V}}$ and $0.4~{\mathrm{V}}$ with $\omega=50~{\mathrm{Hz}}$ and $\Omega=1000~{\mathrm{Hz}}$ and (c) different combinations of $\omega$ and $\Omega$ with $\Omega / \omega=10$ where for the curves $1$, $2$ and $3$ the values of ($\omega,\,\Omega$) are ($50~{\mathrm{Hz}},\,500~{\mathrm{Hz}}$), ($150~{\mathrm{Hz}},\,1500~{\mathrm{Hz}}$) and ($250~{\mathrm{Hz}},\,2500~{\mathrm{Hz}}$) while $f=0.5~{\mathrm{V}}$ and (d) Variation of $Q$ with the parameter $\Omega$ for three fixed values of $\omega$ with $f=0.3~{\mathrm{V}}$ and $g=1.75~{\mathrm{V}}$.}
\label{vrcf5}
\end{figure}

\begin{figure}[th]
\begin{center}
\includegraphics[width=0.55\linewidth]{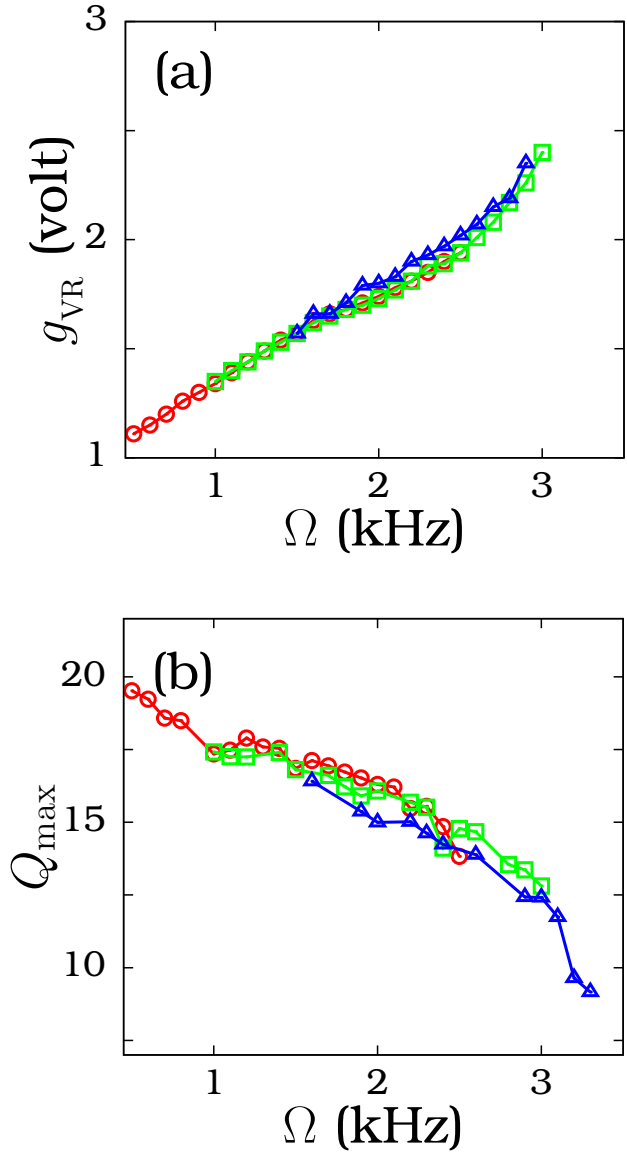}
\end{center}
\caption{Plot of (a) $g_{_{\mathrm{VR}}}$, the critical value of $g$ at which resonance occurs and (b) $Q_{\mathrm{max}}$, the value of $Q$ at $g=g_{\mathrm{VR}}$ versus the high-frequency $\Omega$ of the driving force. The values of the parameters are $f=0.3{\mathrm{V}}$ and $\omega=50~{\mathrm{Hz}}$ (for the symbol circle), $\omega=100~{\mathrm{Hz}}$ (square) and $\omega=150~{\mathrm{Hz}}$ (triangle).}
\label{vrcf6}
\end{figure}

{\noindent We experimentally analyse the influence of the parameters $\omega$, $f$ and $\Omega$ on resonance. Figure \ref{vrcf5} presents the results. In Fig.~\ref{vrcf5}(a) as $\Omega$ increases $g_{_{\mathrm{VR}}}$ also increases but $Q_{\mathrm{max}}$ (the value of $Q$ at resonance) decreases. The width of the bell shape part of the resonance profile increases when $\Omega$ increases. In Fig.~\ref{vrcf5}(b), $g_{_{\mathrm{VR}}}$ decreases while $Q_{\mathrm{max}}$ increases with an increase in $f$. The width of the bell shape part increases for increasing values of  $f$. Figure \ref{vrcf5}(c) shows $Q$ versus $g$ for different set of values ($\omega, \,\Omega$) keeping the ratio $\Omega / \omega$ as $10$. Increase in $\omega$ and $\Omega$ leads to the effect observed in Fig.~\ref{vrcf5}(a). In Fig.~\ref{vrcf5}(d) increase in $\omega$ increases the value of $\Omega_{_{\mathrm{VR}}}$ but decreases the corresponding $Q_{\mathrm{max}}$. Furthermore, we experimentally measure $g_{_{\mathrm{VR}}}$ and $Q_{\mathrm{max}}$ for a range of values of high-frequency $\Omega$ for three different values of $\omega$. Figure \ref{vrcf6} depicts the variation of $g_{_{\mathrm{VR}}}$ and $Q_{\mathrm{max}}$ with $\Omega$. For all the fixed values of $\omega$, $g_{_{\mathrm{VR}}}$ increases while $Q_{\mathrm{max}}$ decreases almost linearly with $\Omega$.}

%%%%%%%%%%%%%%%%%%%%%%%%%%%%%%%%%%%%%%%%%%%%%%%%%%%%%%%%%%%%%%%%%%%%%
%%%%%%%%%%%%%%%%%%%%%%%%%%%%%%%%%%%%%%%%%%%%%%%%%%%%%%%%%%%%%%%%%%%%%
%%%%%%%%%%%%%%%%%%%%%%%%%%%%%%%%%%%%%%%%%%%%%%%%%%%%%%%%%%%%%%%%%%%%%
%%%%%%%%%%%%%%%%%%%%%%%%%%%%%%%%%%%%%%%%%%%%%%%%%%%%%%%%%%%%%%%%%%%%%

\section{System of $n$-coupled Chua's circuits}
In the previous section our focus is on the single Chua's circuit. The present section is devoted to the case of a system of $n$-coupled Chua's circuits with specific emphasis on signal propagation in the presence of a biharmonic external force. Study of coupled nonlinear systems are of great interest in different fields. It has been shown that the response of a nonlinear system can be improved by coupling it into an array of systems. Among the various types of coupling the simple one is the unidirectional linear coupling introduced by Visarath In and his collaborators to induce oscillations in undriven, overdamped and bistable systems \cite{ref37,ref38}. This coupling is found to give rise synchronization \cite{ref39}, propagation of waves of dislocations \cite{ref40}, enhanced signal propagation in coupled overdamped bistable oscillators \cite{ref19} and in coupled maps \cite{ref16} and propagation and annihilation of solitons \cite{ref40, ref34}. Further, it is utilized in electronic sensors and microelectronic circuits \cite{ref42}. We consider a system of unidirectionally coupled $n$-Chua's circuits where only the first circuit is driven by the biharmonic force. We perform a PSpice simulation with $n=75$ units. We show the evidence for improved transmission of low-frequency signal by the combined action of a high-frequency signal and a unidirectional coupling. \\

\begin{figure}[!h]
\begin{center}
\includegraphics[width=0.9\linewidth]{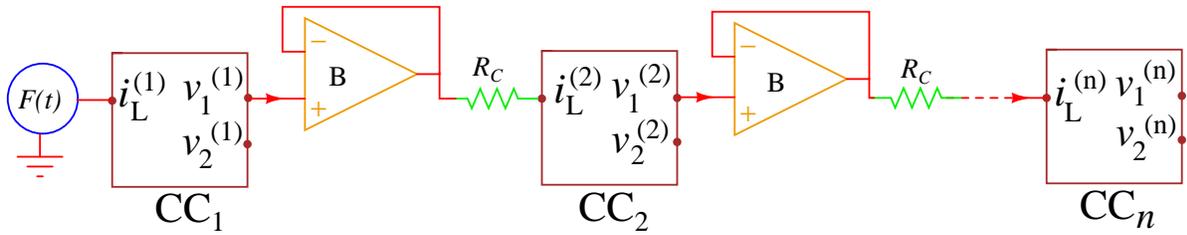}
\end{center}
\caption{The block diagram of a system of $n$ unidirectionally coupled Chua's circuits.  In the first circuit (${\mathrm{CC_1}}$), a biharmonic signal $F(t)$ is connected in series with current ($i_L^{(1)}$) and the remaining circuits are driven by the voltage $v_{1}$ of the previous circuit. Here ${\mathrm{B}}$ represents a buffer circuit and the arrowhead represents the direction of coupling.}
\label{vrcf7}
\end{figure}

The system of $n$-coupled Chua's circuits is shown in Fig.~\ref{vrcf7}. The coupling between the $i$th and ($i+1$)th circuits is made by feeding the voltage across the capacitor $C_1$ of the $i$th circuit to the ($i+1$)th circuit through a buffer as shown in Fig.~\ref{vrcf7}.
The state equations for the coupled circuits shown in Fig. 7 are \cite{ref45}

\begin{subequations}
\label{eq3}
\begin{eqnarray}
C_1\frac{dv_1^{(i)}}{dt}&=&(1/R)\left(v_2^{(i)}-v_1^{(i)}\right)-f\left(v_1^{(i)}\right),\\
C_2\frac{dv_2^{(i)}}{dt}&=&(1/R)\left(v_1^{(i)}-v_2^{(i)}+i_L^{(i)}\right),\\
L\frac{di_L^{(i)}}{dt}&=&-v_2^{(i)}+\delta_i(f\sin\omega t+g\sin\Omega t)+\epsilon_i\left(v_1^{(i-1)}-v_R\right),
\end{eqnarray}
\end{subequations} 

{\noindent where, $\delta_1=1$, $\epsilon_1=0$ and $\delta_i=0$ and $\epsilon_i=1$ for $i=2,3,...,n$ and $v_R=i_{L}^{(i)}R_C$.} 

%%%%%%%%%%%%%%%%%%%%%%%%%%%%%%%%%%%%%%%%%%%%%%%%%%%%%%%%%%%%%%%%%%%%%
%%%%%%%%%%%%%%%%%%%%%%%%%%%%%%%%%%%%%%%%%%%%%%%%%%%%%%%%%%%%%%%%%%%%%
%%%%%%%%%%%%%%%%%%%%%%%%%%%%%%%%%%%%%%%%%%%%%%%%%%%%%%%%%%%%%%%%%%%%%
%%%%%%%%%%%%%%%%%%%%%%%%%%%%%%%%%%%%%%%%%%%%%%%%%%%%%%%%%%%%%%%%%%%%%

{\noindent The buffer circuit makes the coupling as unidirectional. We note that coupled circuits do not form ring since the output of
the last circuit is not fed to the first circuit. The system is a one-way open coupled Chua's circuits. The high input and low output impedances of the buffer ensures that the flow of the signal between $i$th and ($i+1$)th circuits is in forward direction,  that is, from the $i$th circuit to the ($i+1$)th circuit only. The strength of the coupling is characterized by the coupling resistor $R_C$. Ullner et al \cite{ref10} reported propagation of the low-frequency signal in a system of coupled oscillators. They considered excitable oscillators with all the oscillators driven by the high-frequency periodic force. First $100$ oscillators alone driven by the low-frequency force and are uncoupled. The rest of the oscillators are unidirectionally coupled. They performed numerical simulation. In our experimental work, we consider a simple setting of  $n$-coupled circuits.  As shown in Fig. \ref{vrcf7}, the first circuit alone is subject to biharmonic input signal and the coupling is unidirectional. In the systems considered in \cite{ref10}, if the coupling is removed then each oscillator exhibits  oscillatory motion. In contrast to this, in the present case, if the coupling is removed then the first circuit alone exhibits oscillatory motion while the trajectory of all other circuits settle to the stable equilibrium point $X_+$ or $X_-$ depending upon the initial state of the circuit.}\\  

\begin{figure}[!h]
\begin{center}
\includegraphics[width=0.4\linewidth]{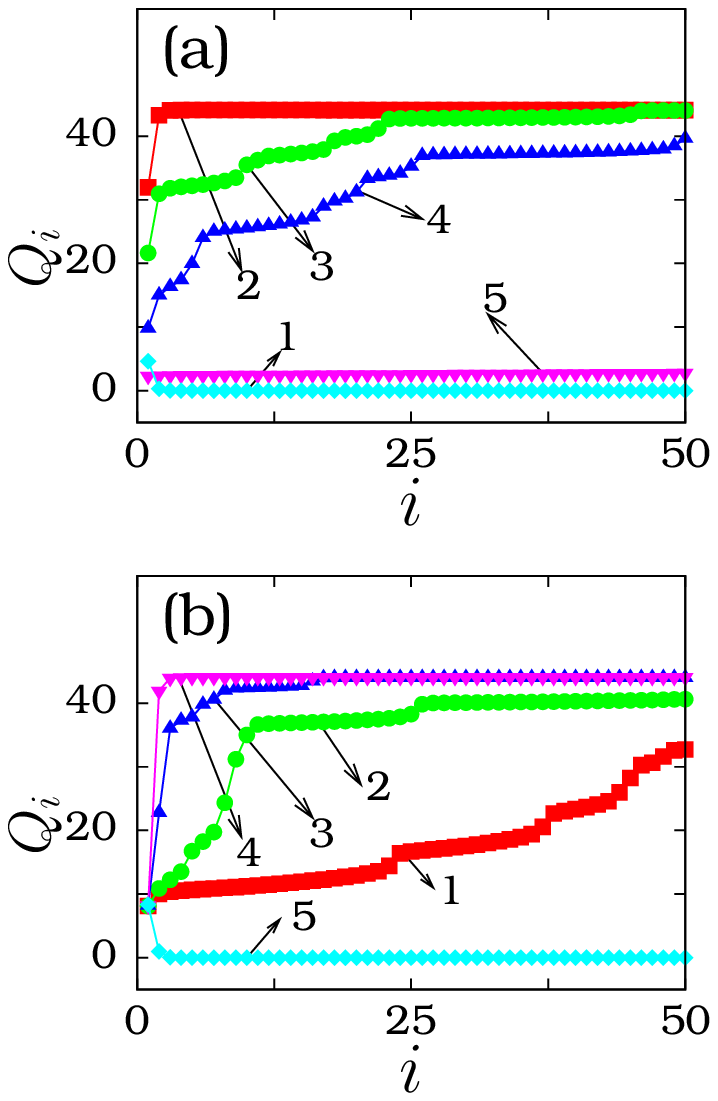}
\end{center}
\caption{(a) $Q_i$ versus $i$ (the number of the Chua's circuit) for five values of $g$ with $R_C=1.0~{\mathrm{k}}\Omega$. For the curves $1-5$ the values of $g$ are $0.6~{\mathrm{V}}$, $0.85~{\mathrm{V}}$, $1.1~{\mathrm{V}}$, $1.15~{\mathrm{V}}$ and $1.6~{\mathrm{V}}$. The values of the other parameters are set as $C_1=10~{\mathrm{nF}}$, $C_2=100~{\mathrm{nF}}$, $L=18~{\mathrm{mH}}$, $R=2.15~{\mathrm{k}}\Omega$, $\omega=100~{\mathrm{Hz}}$, $\Omega=1000~{\mathrm{Hz}}$ and $f=0.3~{\mathrm{V}}$. (b) $Q_i$ versus $i$ for five values of $R_C$ with $g=1.2~{\mathrm{V}}$. The values of $R_C$ for the curves $1-5$ are $1~{\mathrm{k}}\Omega$, $1.2~{\mathrm{k}}\Omega$, $1.4~{\mathrm{k}}\Omega$, $1.8~{\mathrm{k}}\Omega$, and $2~{\mathrm{k}}\Omega$.}
\label{vrcf8}
\end{figure}

We fix the values of the circuit parameters as $C_1=10~{\mathrm{nF}}$, $C_2=100~{\mathrm{nF}}$, $L=18~{\mathrm{mH}}$, $R=2.15~{\mathrm{k}}\Omega$, $\omega=100~{\mathrm{Hz}}$, $\Omega=1~{\mathrm{kHz}}$ and $f=0.3~{\mathrm{V}}$ and treat $R_C$ and $g$ as the control parameters. Figure \ref{vrcf8}(a) presents the variation of $Q_i$ with $i$ (the number of the Chua's circuit) for few values of $g$ where $R_C=1~{\mathrm{k}}\Omega$. For each fixed value of $g$, $Q_i$ varies with $i$ and approaches a limiting value. When $Q_2 > Q_1 (< Q_1)$ then $Q_i$ increases (decreases) with $i$ and reaches a saturation with $Q_{75} > Q_1 (< Q_1)$. The signal propagation through the coupled circuits is termed as undamped when $Q_{75} > Q_1$ otherwise damped. We measure $Q_i$ for $g=1.2~{\mathrm{V}}$ and for five different values of $R_C$. The result is presented in Fig.~\ref{vrcf8}(b). 
%%%%%%%%%%%%%%%%%%%%%%%%%%%%%%%%%%%%%%%%%%%%%%%%%%%%%%%%%%%%%%%%%%%%%
%%%%%%%%%%%%%%%%%%%%%%%%%%%%%%%%%%%%%%%%%%%%%%%%%%%%%%%%%%%%%%%%%%%%%
\newpage
%%%%%%%%%%%%%%%%%%%%%%%%%%%%%%%%%%%%%%%%%%%%%%%%%%%%%%%%%%%%%%%%%%%%%
\begin{figure}[th]
\begin{center}
\includegraphics[width=0.5\linewidth]{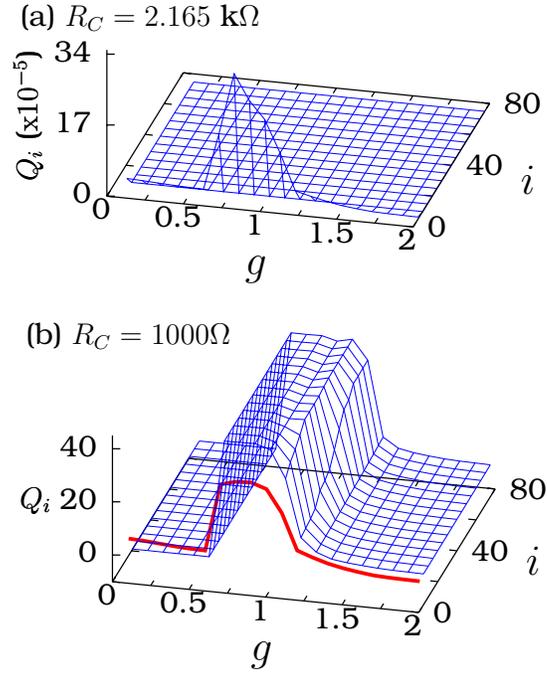}
\end{center}
\caption{$Q_i$ as a function of $i$ and $g$ for two values of $R_C$ illustrating (a) damped propagation of signal (for $R_C=2.165~{\mathrm{k}}\Omega$) and (b) undamped signal propagation (for $R_C=1~{\mathrm{k}}\Omega$) through the unidirectionally coupled Chua's circuits. The thick red curve in (b) represents $Q_1$. In (a) $Q_1$ is not shown because $Q_i$'s with $i>1$ are much lower than $Q_1$.}
\label{vrcf9}
\end{figure}
%%%%%%%%%%%%%%%%%%%%%%%%%%%%%%%%%%%%%%%%%%%%%%%%%%%%%%%%%%%%%%%%%%%%%
%%%%%%%%%%%%%%%%%%%%%%%%%%%%%%%%%%%%%%%%%%%%%%%%%%%%%%%%%%%%%%%%%%%%%
\begin{figure}[b]
\begin{center}
\includegraphics[width=0.5\linewidth]{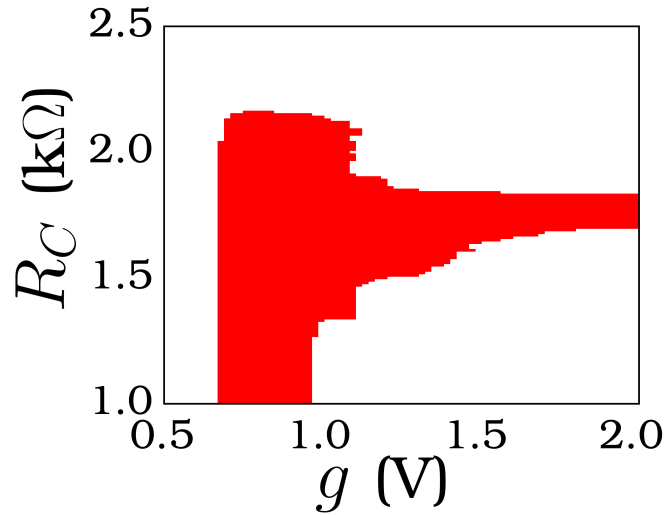}
\end{center}
\caption{Undamped signal propagation (marked by red color) in the ($g-R_C$) parameters space.}
\label{vrcf10}
\end{figure}
%%%%%%%%%%%%%%%%%%%%%%%%%%%%%%%%%%%%%%%%%%%%%%%%%%%%%%%%%%%%%%%%%%%%%
{\noindent Figure \ref{vrcf9} shows $Q_i$ versus $g$ and $i$ for two values of the coupling resistor. For $R_C=2.165~{\mathrm{k}}\Omega$ $Q_i$ versus $g$ (Fig.~\ref{vrcf9}(a)) decays to zero as $i$ increases. This is an example of damped signal propagation. In Fig.~\ref{vrcf9}(b) where $R_C=1~{\mathrm{k}}\Omega$ the signal propagation is undamped $Q_i > Q_1$, $i > 1$ for a range of values of $g$. For $R_C \in [1~{\mathrm{k}}\Omega, \, 2.5~{\mathrm{k}}\Omega]$ and $g \in[0.5~{\mathrm{V}}, \, 2.0~{\mathrm{V}}]$ we experimentally identify the regions for which undamped signal propagation occurs. Figure \ref{vrcf10} displays the result. Only for certain set of values of $R_C$ and $g$ undamped signal propagation occurs. $Q_{75}=0$ for (i) all values of $R_C$ if $g<0.68~{\mathrm{V}}$ and (ii) all values of $g$ if $R_C \geq 2.165~{\mathrm{k}}\Omega$. Figure \ref{vrcf11} presents another interesting result. In this figure we have plotted $v_1$ of the $i$th circuit versus $t$ for four values of $i$. $v_1$ is periodic with period $T=1 / \omega$. The $v_1$ of the first circuit ($i=1$) is modulated by the high-frequency drive. Since $\Omega/\omega=10$, $v_1$ has ten peaks over one period. The high-frequency oscillation weakens as the number of the circuit $i$ increases as seen clearly in Figs.~\ref{vrcf11}(b) and (c) for $i=5$ and $i=15$ respectively. For sufficiently large $i$ the amplitude modulation of $v_1$ disappears. Moreover, the output signal appears as a rectangular pulse (Fig.~\ref{vrcf11}(d)), of low-frequency $\omega$. We wish to emphasize that in the coupled Chua's circuits the biharmonic input signal is applied only to the first circuit. Essentially, the unidirectional coupling serves as a low-pass filter by weakening the propagation of the high-frequency component while enhancing the low-frequency component for a range of values $g$ and $R_C$ of the circuits.
%%%%%%%%%%%%%%%%%%%%%%%%%%%%%%%%%%%%%%%%%%%%%%%%%%%%%%%%%%%%%%%%%%%%%
\begin{figure}[ht]
\begin{center}
\includegraphics[width=0.53\linewidth]{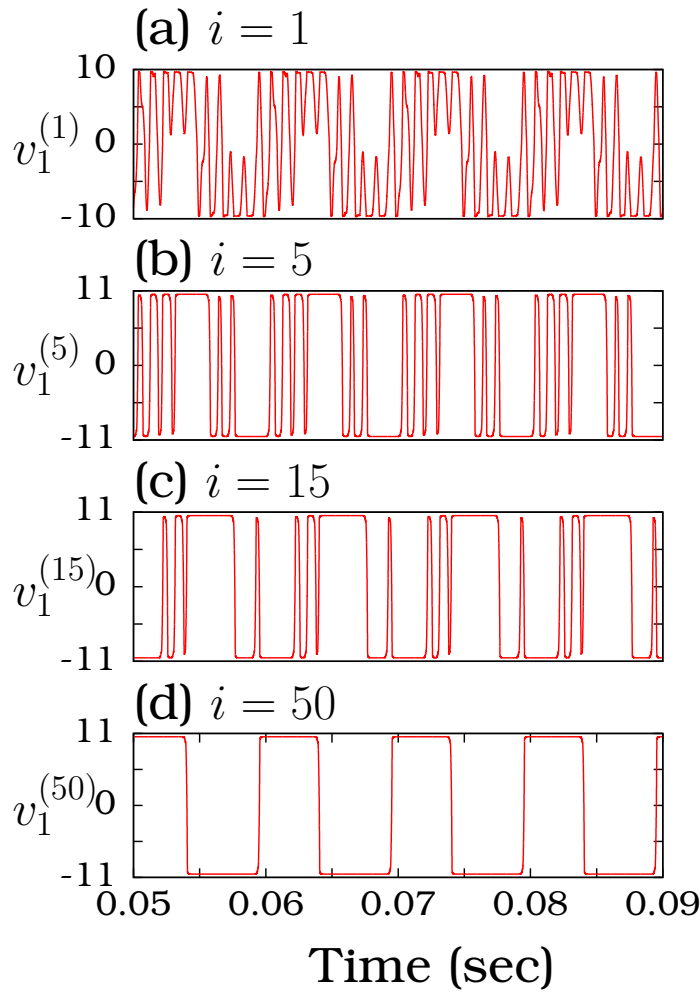}
\end{center}
\caption{Evolution of $v_1^{(i)}$ with time at four different nodes denoted as $i$ where $R=2.15~{\mathrm{k}}\Omega$, $R_C=1~{\mathrm{k}}\Omega$ and $g=1.1~{\mathrm{V}}$. Notice the suppression of high-frequency oscillations as $i$ increases.}
\label{vrcf11}
\end{figure}
%%%%%%%%%%%%%%%%%%%%%%%%%%%%%%%%%%%%%%%%%%%%%%%%%%%%%%%%%%%%%%%%%%%%%

Finally, we consider the $n$-coupled Chua's circuit system with bidirectional coupling.  In this case, the output of the $i$th circuit is fed as an input to the ($i+1$)th circuit and that of the circuit ($i+1$) is fed as an input to the $i$th circuit. The strength of the coupling is characterised by the coupling resistor $R_C$. We keep the values of the circuit parameters as chosen for unidirectional coupling and treat $R_C$ and $g$ as the control parameters. Figure \ref{vrcf12} presents the variation of response amplitude $Q_i$ with $g$ and $i$ for $R_C=1000\Omega$. It clearly shows that the signal propagation is damped over the chain. We identified the nature of signal propagation for a wide range of values of $g$ and $R_C$. Undamped signal propagation observed in the unidirectionally coupled circuits is not observed in the bidirectionally coupled circuits.}\\
%%%%%%%%%%%%%%%%%%%%%%%%%%%%%%%%%%%%%%%%%%%%%%%%%%%%%%%%%%%%%%%%%%%%%
\begin{figure}[t]
\begin{center}
\includegraphics[width=0.55\linewidth]{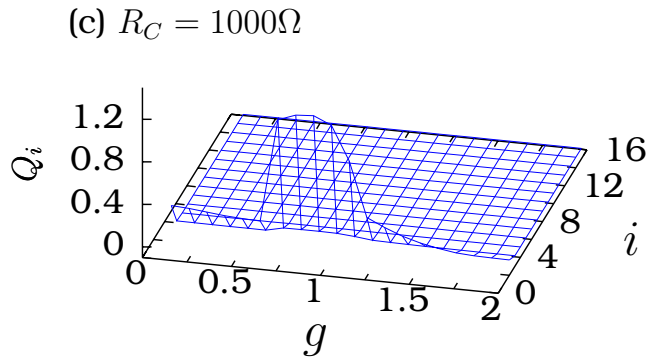}
\end{center}
\caption{$Q_i$ as a function of $g$ and $i$ for $R_C=1000\Omega$. For clarity $Q_1$ is not shown because $Q_1(g)$ is $\gg Q_i$ for $i=2,3...$.}
\label{vrcf12}
\end{figure}
%%%%%%%%%%%%%%%%%%%%%%%%%%%%%%%%%%%%%%%%%%%%%%%%%%%%%%%%%%%%%%%%%%%%%

\section{Conclusion}

In the present work, we have considered one of the the most widely investigated nonlinear circuits, namely the Chua's circuit which is capable of displaying a variety of complex dynamics. Here, we have shown that the Chua's circuit can also be used for weak signal detection and amplification through the vibrational resonance phenomenon. Resonance occurs when the output signal switches between two stable equilibrium states periodically with a period $T/2$ where $T$ is the period of the low-frequency input signal. \\

The PSpice simulation study of a system of $n$-coupled Chua's circuits reveals  undamped signal propagation for a range of values of the amplitude $g$ of the high-frequency input signal and the coupling parameter $R_C$. Another interesting result is the suppression of the high-frequency component in distant circuits while maintaining an enhanced signal propagation of low-frequency signal. The $Q_i$ either decays to zero or approaches a nonzero constant values with increase in $i$. For most of the parametric choices considered in the present work, $Q_i$ attained a stationary value for $i>50$. Therefore we have considered 75 coupled circuits. Study of stochastic, coherence and vibrational resonances in variants of Chua's circuits  such as switch controlled Chua's circuit and multi-scroll Chua's circuit would bring out practical applications of these circuits in both weak signal detection (ac as well as dc) and output signal amplification. \\

A theoretical method has been developed to investigate vibrational resonance in nonlinear oscillators with polynomial potentials \cite{ref1,ref2,ref8}. In this approach one can obtain linear system for the low-frequency component of the solution. Since, the equation of motion of the Chua's circuit is piecewise linear it is very difficult to separately write the equations of motion for the slow and fast components (with frequencies $\omega$ and $\Omega$ respectively).

\section*{Acknowledgments}

R.J. is supported by the University Grants Commission, India in the form of Research Fellowship in Science for Meritorious Students. The work of K.T. forms a part of a Department of Science and Technology, Government of India sponsored project grant no. SR/S2/HEP-015/2010. M.A.F.S. acknowledges financial support from the Spanish Ministry of Science and Innovation under project no. FIS2009-09898.


\begin{thebibliography}{50}
\bibitem[Baltan\'as {\it et al.}(2003)]{ref7}
Baltan\'as, J. P., L\'opez, L., Blechman, I. I., Landa, P. S., Zaikin, A., Kurths, J., \& Sanju{\'a}n, M. A. F. [2003] ``Experimental evidence, numerics and theory of vibrational resonance in bistable systems," {\it Phys. Rev. E} {\bf{67}}, 066119.
%
\bibitem[Bhowmick {\it et al.}(2012)]{ref26}
Bhowmick, S. K., Amritkar, R. E., \& Dana, S. K. [2012] ``Experimental evidence for synchronization of the time-varying dynamic network," {\it Chaos} {\bf{22}}, 023105.
%
\bibitem[Blekhman \& Landa(2004)]{ref2}
Blekhman, I. I., \& Landa, P. S. [2004] ``Conjugate resonances and bifurcations in nonlinear systems under biharmonic excitation," {\it Int. J. Non-Linear Mech.} {\bf{39}}, 421--426.
%
\bibitem[Breen {\it et al.}(2011)]{ref34}
Breen, B. J., Doud, A. B., Grimm, J. R., Tanasse, A. H., Tanasse, S. J., Lindner, J. F., \& Maxted, K. J. [2011] ``Electronic and mechanical realization of one-way coupling in one and two dimensions," {\it Phys. Rev. E} {\bf{83}}, 037601.
%
\bibitem[Buscarino {\it et al.}(2009)]{ref20}
Buscarino, A., Fortuna, L., \& Frasca, M. [2009] ``Jump resonance in driven Chua's circuit," {\it Int. J. Bifur. Chaos} {\bf{19}}, 2557--2562.
%
\bibitem[Chakraborty \& Dana(2010)]{ref22}
Chakraborty, S., \& Dana, S. K. [2010] ``Shi{\l'}nikov chaos and mixed mode oscillations in Chua circuit," {\it Chaos} {\bf{20}} 023107.
%
\bibitem[Chen \& Ueta(2002)]{ref46}
Chen, G. \& Ueta, T. [2002] {\it Chaos in Circuits and Systems} (world Scientific, Singapore).
%
\bibitem[Chen \& Duan(2008)]{ref28}
Chen, G. \& Duan, Z. [2008] ``Network synchronizability analysis: A graph theoretic approach," {\it Chaos} {\bf{18}}, 037102.
%
\bibitem[Chizhevsky \& Giacomelli(2006)]{ref14}
Chizhevsky, V. N. \& Giacomelli, G. [2006] ``Experimental and theoretical study of vibrational resonance in a bistable system with asymmetry," {\it Phys. Rev. E} {\bf{73}}, 022103.
%
\bibitem[Chizhevsky(2008)]{ref8}
Chizhevsky, V. N. [2008] ``Analytical study of vibrational resonance in an overdamped bistable oscillator," {\it Int. J. Bifur. Chaos} {\bf{18}}, 1767--1774.
%
\bibitem [Chizhevsky \& Giacomelli(2008)]{ref15}
Chizhevsky, V. N. \& Giacomelli, G. [2008] ``Vibrational resonance and detection of aperiodic binary signal," {\it Phys. Rev. E} {\bf{77}}, 051126.
% 
\bibitem[Deng {\it et al.}(2010)]{ref18}
Deng, B., Wang, J., Wei, X., Tsang, K. M., \& Chan, W. L. [2010] ``Vibrational resonance in neuronal populations," {\it Chaos} {\bf{20}}, 013113.
%
\bibitem[Daza {\it et al.}(2013)]{ref13}
Daza, A., Wagemakers, A., Rajasekar, S. \& Sanju{\'a}n, M. A. F. [2013] ``Vibrational resonance in time-delayed genetic toggle switch," {\it Commun. Nonlinear Sci. Numer. Simulat.} \textbf{18}, 411-416.
%
\bibitem[Fortuna {\it et al.}(2009)]{ref44}
Fortuna, L.,  Frasca, M., \& Xibilia, M. G. {\it Chua's circuit implementations:Yesterday, Today and Tomorrow} (World Scientific, Singapore).
%
\bibitem[Gammaitoni {\it et al.}(1998)]{ref3}
Gammaitoni, L., Hanggi, P., Jung, P., \& Marachesoni, F. [1998] ``Stochastic resonance," {\it Rev. Mod. Phys.} {\bf {70}}, 223--287.
%
\bibitem[Gitterman(2001)]{ref6}
Gitterman, M. [2001] ``Bistable oscillator driven by two periodic fields," {\it J. Phys. A: Math. Gen.} {\bf{34}}, L355--L357.
%
\bibitem[Gomes {\it et al.}(2012)]{ref24}
Gomes, I., Vermelho, M. V. D., \& Lyra, M. L. [2012] ``Ghost resonance in the Chaotic Chua's circuit," {\it Phys. Rev. E} {\bf{85}}, 056201.
%
\bibitem[Heo {\it et al.}(2012)]{ref32}
Heo, Y. S., Jung, J. W., Kim, J. M., Jo, M. K., \& Song, H. J. [2012] ``Chaotic dynamics of a Chua's system with voltage controllability," {\it J. Korean Phys. Soc.} {\bf{60}}, 1140--1144.
%
\bibitem[In {\it et al.}(2003a)]{ref37}
In, V., Bulsara, A. R., Palacios, A., Longhini, P., Kho, A., \& Neff, J. D. [2003a] ``Coupling-induced oscillations in overdamped bistable system," {\it Phys. Rev. E} {\bf{68}}, 045102(R).
%
\bibitem[In {\it et al.}(2003b)]{ref38}
In, V., Kho, A., Neff, J. D., Palacios, A., Longhini, P., \& Meadows, B. K. [2003b] ``Experimental observation of multifrequency patterns in array of coupled nonlinear oscillators," {\it Phys. Rev. Lett.} {\bf{91}}, 244101.
%
\bibitem[In {\it et al.}(2005)]{ref39}
In, V., Bulsara, A. R., Palacios, A., Longhini, P., \& Kho, A. [2005] ``A bistable micro electronic circuit for sensing extremely low electric field," {\it Phys. Rev. E} {\bf {72}}, 045104(R).
%
\bibitem[Jeevarathinam {\it et al.}(2011)]{ref12}
Jeevarathinam, C., Rajasekar, S., \& Sanju{\'a}n, M. A. F. [2011]  ``Theory and numerics of vibrational resonance in Duffing oscillators with time-delayed feedback," {\it Phys. Rev. E} {\bf{83}}, 066205.
%
\bibitem[Jeyakumari {\it et al.}(2009)]{ref5}
Jeyakumari, S., Chinnathambi, V., Rajasekar, S. \& Sanju{\'a}n, M. A. F. [2009] ``Single and multiple vibrational resonance in a quintic oscillator with monostable potentials," {\it Phys. Rev. E} {\bf{80}}, 046608.
%
\bibitem[Kapitaniak {\it et al.}(1994)]{ref45}
Kapitaniak, T., Chua, L.O., \& Zhong, G. ``Experimental hyperchaos in coupled Chua's circuits." {\it IEEE Trans. circuits and syst. -I:}, {\bf {41}}, 499--503.
%
\bibitem[Koofigar {\it et al.}(2011)]{ref29}
Koofigar, H. R., Sheikholeslam, F., \& Hosseinnia, S. [2011] ``Robust adaptive synchronization for a general class of uncertain chaotic systems with application to Chua's circuit," {\it Chaos} {\bf{21}}, 043134.
%
\bibitem[Landa \& McClintock(2000)]{ref1}
Landa, P. S., \& McClintock, P. V. E. [2000] ``Vibrational resonance," {\it J. Phys. A: Math. Gen.} {\bf{33}}, L433--L438.
%
\bibitem[Lakshmanan \& Murali(1996)]{ref43} 
Lakshmanan, M., \& Murali, K. {\it Chaos in Nonlinear Oscillators: Controlling and Synchronization} (World Scientific, Singapore).
%
\bibitem[Lindner \&  Bulsara(2006)]{ref40}
Lindner, J. F., \&  Bulsara, A. R. [2006] ``One-way coupling enables noise-mediated spatiotemporal patterns in a media of otherwise quiescent multistable elements," {\it Phys. Rev. E} {\bf{74}}, 020105(R).
%
\bibitem[Lindner {\it et al.}(2008)]{ref41}
Lindner, J. F., Patton, K. M., Odenthal, P. M., Gallagher, J. C., \& Breen, B. J. [2008] ``Experimental observation of soliton propagation and annihilation in a hydrodynamical array of one-way coupled oscillators," {\it Phys. Rev. E} {\bf{78}}, 066604.
%
\bibitem[McDonnel {\it et al.}(2008)]{ref4}
McDonnel, M. D., Stokes, N. G., Pearce, C. E. M., \& Abbott, D. [2008] {\it Stochastic Resonance} (Cambridge University Press, Cambridge).
%  
\bibitem[Nurujjaman {\it et al.}(2012)]{ref30}
Nurujjaman, Md., Shivamurthy, S., Apte, A., \& Singla, T. [2012] ``Effect of discrete time observations on synchronization in Chua model and application to data assimilation," {\it Chaos} {\bf{22}}, 023125.
%
\bibitem[Ozden {\it et al.}(2004)]{ref31}
Ozden, I., Venkataramani, S., Long, M. A., Connors, B. W., \& Nurmikko, A. V. [2004] ``Strong coupling of nonlinear electronic and biological oscillators: Reaching the ``Amplitude Death" regime," {\it Phys. Rev. Lett.} {\bf{93}}, 158102.
%
\bibitem[Petras(2010)]{ref25}
Petras, I. [2010] ``Fractional order memristor based Chua's circuit," {\it IEEE Trans. Circuits Syst. II: Express Briefs} {\bf{57}}, 975--979.
%
\bibitem[Rajasekar {\it et al.}(2011)]{ref9}
Rajasekar, S., Abirami, K., \& Sanju{\'a}n, M. A. F. [2011] ``Novel vibrational resonance in multistable systems," {\it Chaos} {\bf{21}}, 033106.
%
\bibitem[Rajasekar {\it et al.}(2012)]{ref16}
Rajasekar, S., Used, J., Wagemakers, A., \& Sanju{\'a}n, M. A. F. [2012] ``Vibrational resonance in biological nonlinear maps," {\it Commun. Nonlinear Sci. Numer. Simulat.} {\bf{17}}, 3435--3445.
%
\bibitem[Ramirez-Avila \& Gallas(2010)]{ref21}
Ramirez-Avila, G. M., \& Gallas, J. A. [2010] ``How similar is the performance of the cubic and piecewise-linear circuits of Chua," {\it Phys. Lett. A} {\bf{375}}, 143--148.
%
\bibitem[Razavi(2008)]{ref42}
Razavi, B. [2008] {\it{Fundamentals of Microelectronics}} (John-Wiley, New Delhi).
%
\bibitem[Roberts \& Sedra(1995)]{ref36} 
Roberts, G. W. \& Sedra, A. S. [1995] {\it{SPICE}} (Oxford University Press, New York).
%
\bibitem[Singla {\it et al.}(2011)]{ref23}
Singla, T., Pawar, N., \& Parmananda, P. [2011] ``Exploring the dynamics of conjugate coupled Chua's circuits: Simulations and experiments," {\it Phys. Rev. E} {\bf{83}}, 026210.
%
\bibitem[Tuinenga(1995)]{ref35}
Tuinenga, P. W. [1995] {\it{SPICE-A guide to circuit simulation and analysis using PSpice}} (Prentice Hall, New Jersey).
%
\bibitem[Ullner {\it et al.}(2003)]{ref10}
Ullner, E., Zaikin, A., Garcia-Ojalvo, J., Bascones, R. \& Kurths, J. [2003] ``Vibrational resonance and vibrational propagation in excitable systems," {\it Phys. Lett. A} {\bf{312}}, 348--354.
%
\bibitem[Wai {\it et al.}(2011)]{ref27}
Wai, R. J., Lin, Y. W.,  \& Yang, H. C. [2011] ``Experimental verification of total sliding-mode control for Chua's chaotic circuit," {\it IET Circuits Devices Syst.} {\bf{5}}, 451--461.
%
\bibitem[Yang \& Liu(2010)]{ref11}
Yang, J. H. \& Liu, X. B. [2010] ``Delay induces quasi-periodic vibrational resonance," {\it J. Phys. A: Math. Theor.} {\bf{43}}, 122001.
%
\bibitem[Yao \& Zhan(2010)]{ref19}
Yao, C., \& Zhan, M. [2010] ``Signal transmission by vibrational resonance in one-way coupled bistable systems," {\it Phys. Rev. E} {\bf{81}}, 061129.
%
\bibitem[Yu {\it et al.}(2011)]{ref17}
Yu, H., Wang, J., Liu, C., Deng, B., \& Wei, X. [2011] ``Vibrational resonance in excitable neuronal systems," {\it Chaos} {\bf{21}}, 043101.
%
\bibitem[Zhou \& Song(2012)]{ref33}
Zhou, J. C., \& Song, H. J. [2012] ``Chaotic dynamics of a  three-phase clock-driven oscillator with dual voltage controllability," {\it J. Korean Phys. Soc.} {\bf{61}}, 1303--1307.
%
\end{thebibliography}
\end{document}